\documentclass[12pt,preprint]{aastex}
\usepackage{apjfonts}

\makeatletter

\begin{document}

\shortauthors{Lahuis et al.}
\shorttitle{Molecular Starburst Fingerprints in (U)LIRG Nuclei}

\title{Infrared Molecular Starburst Fingerprints in Deeply Obscured (U)LIRG Nuclei}
\author{F.~Lahuis\altaffilmark{1,2},
        H.~W.~W.~Spoon\altaffilmark{3,4},
        A.~G.~G.~M.~Tielens\altaffilmark{5},
        S.~D.~Doty\altaffilmark{6},
        L.~Armus\altaffilmark{7},
        V.~Charmandaris\altaffilmark{8,9}, 
        J.~R.~Houck\altaffilmark{3},
        P.~St\"auber\altaffilmark{10}, \&
        E.~F.~van~Dishoeck\altaffilmark{1}
       }

\altaffiltext{1}{Leiden Observatory, P.O. Box 9513, 2300 RA Leiden, 
                 The Netherlands}
\altaffiltext{2}{SRON Netherlands Institute for Space Research, 
                 P.O. Box 800, 9700 AV Groningen, The Netherlands}
\email{F.Lahuis@sron.rug.nl}
\altaffiltext{3}{Cornell University, Astronomy Department, Ithaca, NY 14853}
\altaffiltext{4}{{\it Spitzer} Fellow}
\altaffiltext{5}{MS 245-3, NASA Ames Research Center, Moffett Field, CA 94035}
\altaffiltext{6}{Department of Physics and Astronomy, Denison University,
                 Granville, OH 43023, USA}
\altaffiltext{7}{Caltech, {\it Spitzer} Science Center, MS 220-6, Pasadena, CA 91125}
\altaffiltext{8}{Department of Physics, University of Crete,
                 GR-71003, Heraklion, Greece}
\altaffiltext{9}{IESL / Foundation for Research and Technology-
                 Hellas, PO Box 1527, GR-71110, Heraklion, Greece, and 
                 Chercheur Associ\'e, Observatoire de Paris, F-75014,  Paris, France}
\altaffiltext{10}{Institute of Astronomy, ETH-Zentrum, 8092 Zurich, Switzerland}

\setcounter{footnote}{10}
\begin{abstract}
High resolution spectra of the {\it Spitzer} Space Telescope 
show vibration-rotation 
absorption bands of gaseous C$_2$H$_2$, HCN, and CO$_2$ molecules toward a 
sample of deeply obscured (U)LIRG nuclei. The observed bands reveal the 
presence of dense ($n\gtrsim 10^7\,\mathrm{cm}^{-3}$), warm 
($T_\mathrm{ex}=200 - 700\,\mathrm{K}$) molecular gas with high column 
densities of these molecules ranging from a few 
$10^{15} - 10^{17}\,\mathrm{cm}^{-2}$. 
Abundances relative to H$_2$, inferred from the silicate optical depth, range 
from $\sim10^{-7}$ to $10^{-6}$ and show no correlation with temperature.
Theoretical studies show that the high abundances of both C$_2$H$_2$ 
and HCN exclude a X-ray dominated region (XDR) associated with the 
toroid surrounding an AGN as the origin of this dense warm molecular gas. 
Galactic massive protostars in the so-called Hot Core phase have
similar physical characteristics with comparable high abundances of 
C$_2$H$_2$, HCN, and CO$_2$ in the hot phase. However, the abundances of 
C$_2$H$_2$ and  HCN  and the C$_2$H$_2$/CO$_2$ and HCN/CO$_2$ ratios
are much higher toward the (U)LIRGs in the cooler 
($T_\mathrm{ex}\lesssim 400$\,K) phase.
We suggest that the warm dense molecular gas 
revealed by the mid-IR absorption lines is associated with a phase 
of deeply embedded star formation where the extreme pressures and 
densities of the nuclear starburst environment have inhibited the 
expansion of H{\sc ii} regions and the global disruption of the star 
forming molecular cloud cores, and `trapped' the star formation 
process in an `extended' Hot Core phase.
\end{abstract}


\section{Introduction}
\label{sec:introduction} 
One of the holy grails in the study of 
luminous and ultraluminous infrared galaxies (LIRGs and ULIRGs)
is to elucidate the true nature of the central energy source. 
(U)LIRGs emit nearly all their energy in the mid and far-infrared 
part of the spectrum. LIRGs have a luminosity 
$L_{8-1000\mu\mathrm{m}} > 10^{10} L_\odot$ and ULIRGs 
$L_{8-1000\mu\mathrm{m}} > 10^{12} L_\odot$, equal to the power output of quasars.
(U)LIRGs are generally found in interacting and merging systems 
\citep*[e.g][]{armus87,sanders88a,murphy96}.
During the merger large amounts of gas and dust are concentrated toward 
the nucleus \citep*[e.g.][]{mihos96}, fueling a massive starburst and 
possibly a massive black hole (an AGN).
(U)LIRGs are without doubt the most spectacular sites of star formation
in the Universe and if a substantial part of their energy originates from
AGN activity, it would show the AGN in its very earliest deeply 
enshrouded phase.

The sensitive Infrared Spectrograph (IRS) \citep{houck04} on board the 
{\it Spitzer} Space Telescope \citep*{werner04} has revealed the richness, 
complexity and diversity of the mid-infrared spectra toward a sample of
deeply obscured (U)LIRGs \citep*[][]{armus04,armus06a,armus06b,spoon06a,spoon06b}; 
see \citet*[][]{genzel00} for a review of results from the 
Infrared Space Observatory.
The general characteristic of these spectra is the presence of deep, broad 
absorption of amorphous silicates, centered at 10 and 18 $\mu$m. In addition, the 
spectra show a large variety in absorption features of crystalline 
silicates, aliphatic hydrocarbons, water ice, and gas phase bands of 
hot CO \citep*[][]{spoon04,spoon06c}. PAH emission bands are generally weak 
and in some cases absent.
Absorption bands of more volatile ices (e.g. CO or CO$_2$), commonly 
detected in Galactic dense molecular clouds, are generally absent or 
very weak.

The very compact nuclei of deeply obscured (U)LIRGs are packed with gas.
Molecular gas has been observed at millimeter wavelengths 
--- through its low lying pure rotational transitions --- in (U)LIRG 
nuclei \citep*[e.g.][]{solomon97,downes98}, see \citet*{aalto05} for 
a recent overview. In recent years millimeter emission lines of HCN 
and HCO$^+$ have been observed revealing the presence of relatively 
dense ($n_\mathrm{H}\sim10^5\,\mathrm{cm}^{-3}$) molecular gas 
\citep*[][]{imanishi06,kohno05}. 
However, the analysis of the millimeter emission lines is complicated 
as a result of beam dilution which strongly depends on the molecules 
and transitions observed.

The IRS Short-High spectrometer (IRS-SH) on board the {\it Spitzer} 
Space Telescope allows for the first time the direct study of the 
very dense and warm molecular gas in the central regions of deeply 
obscured luminous and ultraluminous infrared galaxies through IR pencil-beam 
line-of-sight absorption spectroscopy against the continuum of individual 
nuclei or unresolved%
\footnote{The IRS-SH slit width is 4.7", equal to the size of the PSF at $19.5\mu$m.}
double nuclei.
In particular, the mid-infrared vibration-rotation bands of C$_2$H$_2$ 
and HCN uniquely trace warm ($100<T<1000$\,K) and dense 
($n_\mathrm{H}>10^7\,\mathrm{cm}^{-3}$) molecular gas. 
These bands have previously been detected primarily toward Galactic
massive protostars \citep*[][]{lacy89,evans91,lahuis00}.
They may prove to be a new strong tool to probe the heating sources 
of deeply obscured (U)LIRG nuclei (starburst or AGN activity).

\section{Observations}
The observations presented in this paper are part of the {\it Spitzer}
IRS observing programs (PID) 105 (IRS GTO ULIRG program, J.R.\,Houck PI),
96 (IRS GTO program on nearby AGN, J.R.\,Houck PI) and 1096
(Director's Discretionary Time (DDT) proposal, H.W.W.\,Spoon PI).
The IRS GTO ULIRG sample comprises $\sim$100 ULIRGs in the redshift
range 0.02 $< z <$ 0.93, selected primarily from the IRAS
2-Jy sample \citep{strauss92}, IRAS 1-Jy sample \citep{kim98}
and the FIRST/IRAS radio-far-IR sample \citep{stanford00}.
The samples of PIDs 96 and 1096 contain 3 additional ULIRGs
(IRAS\,04384--4848, IRAS\,03000--2719 and IRAS\,02113--2937)
and 3 additional LIRGs (IRAS 02530+0211, IC\,860 and NGC\,4418).
For all sources in this combined sample low-resolution spectra
($R=\lambda/\Delta\lambda\sim100$) have been obtained, while high 
resolution spectra ($R=600$) have been taken only for the 
brighter half of the sample.

All high resolution spectra in this sample have been investigated
for the presence of vibration-rotation absorption bands of
C$_2$H$_2$ (13.7\,$\mu$m), HCN (14.02\,$\mu$m) and CO$_2$ (15.0\,$\mu$m)
against the nuclear continuum. 
Fifteen of the sources listed in Table \ref{tab:observations} show 
absorption due to (some of) these species.
Because of the low signal to noise ratio of the spectra and/or the low H-column 
densities (as derived from the 9.7 um silicate optical depth; see Section
\ref{sec:abundances}), the derived upper limits on the molecular column 
densities for all other sources do not place very stringent constraints
on the C2H2 and HCN abundances. 
To illustrate this, four sources without positive detections with both a 
moderate-to-large hydrogen column density and good signal to noise 
spectra, are included in Tables \ref{tab:observations} and \ref{tab:analysis}. 
The upper limits on the C$_2$H$_2$ and HCN abundances fall within the 
range of derived abundances toward the fifteen sources.

Data reduction started from crosstalk corrected echelle images
using S12 and S13 {\it Spitzer} archive data. Processing was done using
the c2d analysis pipeline \citep{kessler06,lahuis06}.
It includes echelle image analysis (a.o. bad-pixel detection, image arithmic, 
and optimal spectral extraction), 
defringing of the extracted spectra \citep{lahuis03}, and spectral analysis
(see Section \ref{sec:mol_analysis}).
The optimal spectral extraction uses an analytical source profile defined
and calibrated using a suite of calibrator stars. Calibration is
done using MARCS stellar models provided by the {\it Spitzer} Science Center
\citep{decin04}. The optimal spectral extraction employs a combined
source and local sky fit. This provides wavelength dependent sky estimates
and allows to discriminate between resolved and unresolved spectral
features.

\begin{figure*}
\plotone{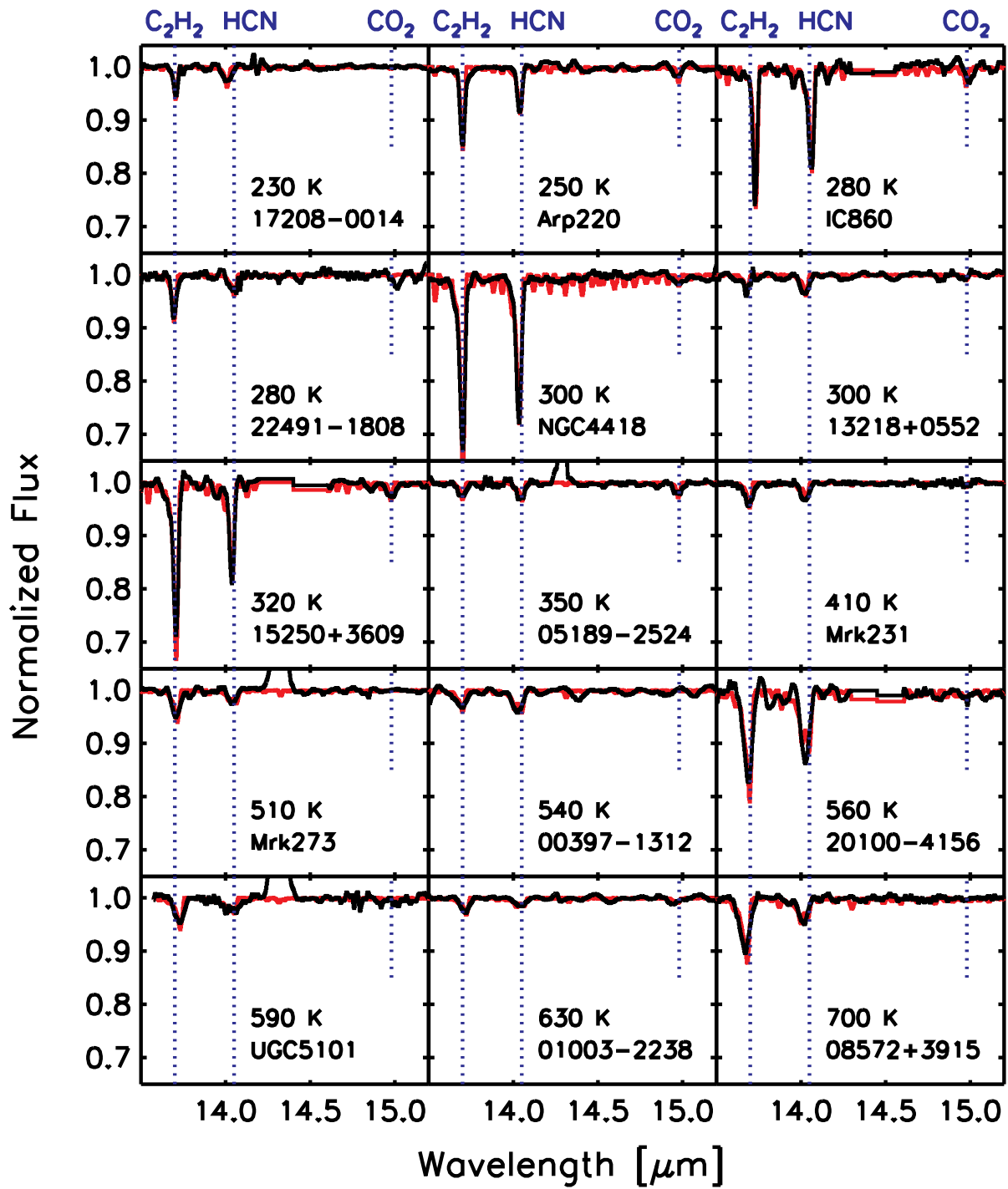}
\caption{\label{fig:all_ulirgs}%
Continuum divided {\it Spitzer}-IRS spectra of a sample of (U)LIRGs showing 
the absorption bands of C$_2$H$_2$ and HCN and some of CO$_2$. 
Plotted in red are best-fit synthetic spectra
assuming a single excitation temperature for all three molecules.
All spectra have been shifted to the rest wavelengths. 
}
\end{figure*}

Figure \ref{fig:all_ulirgs} shows the continuum divided, sky and
redshift-corrected spectra from $13.5 - 15.5\,\mu$m covering the 
absorption bands of C$_2$H$_2$, HCN and CO$_2$. 
Plotted in red are the best-fit single temperature synthetic spectra 
(see Section \ref{sec:results}). Indicated with the dotted verticals 
are the positions of the ground-state $Q$-branch transitions of the three 
molecules. For IRAS\,15250+3609, IRAS\,20100--4156, and IC\,860 a small 
section of the spectrum in between HCN and CO$_2$ is affected by 
artifacts in inter-order sections of the spectrum, and these sections 
have been clipped from the presented spectra.

\section{Analysis}
\label{sec:results}

\subsection{Molecular analysis}
\label{sec:mol_analysis}
The {\it Spitzer} spectra unambiguously reveal the presence of the 
$Q$-branch transitions of C$_2$H$_2$ \hbox{$\nu_5=1-0$}, HCN $\nu_2=1-0$ 
and CO$_2$ $\nu_2=1-0$, each of which blends into a
"broad'' absorption feature. The corresponding $P$- and $R$-branch
transitions of these species are difficult to observe with the
{\it Spitzer}-IRS due to spectral dilution at the IRS resolving power and
the presence of instrumental fringe residuals with frequencies and 
amplitudes close to those of the $P$- and $R$-branch lines.
The $Q$-branch transitions are analyzed using a pure absorption model
assuming local thermodynamic equilibrium (LTE) excitation of the
levels at a single temperature. The adopted method is described in
detail in \citet{lahuis00} and \citet*{boonman03}, which includes
references to the molecular parameters and data used in the model.
The main fit parameters are the excitation temperature and 
column density along the line of sight for a given intrinsic line 
width, defined by the Doppler $b$-value. 
It is assumed that the absorbing molecules have a covering factor
of unity of the continuum, i.e. the mid-IR continuum is composed solely
by the regions toward which the molecular absorption features arise. 
A covering factor less than unity, i.e. larger continuum emitting
regions, increases the true optical depth of the absorption features 
resulting in higher column densities and possibly lower temperatures 
(when lowest excitation lines saturate at large column densities).

The derived excitation parameters do not strongly depend on the
exact value of $b$. Only for low values 
of $b$ ($\lesssim 5$\,km\,s$^{-1}$) will saturation result in 
an underestimate of the column density for the 
sources with the largest column densities. Such low $b$ values are
expected for quiescent gas where thermal broadening dominates.
However, non-thermal broadening will likely dominate in the dense 
energetic ISM of the galactic nuclei and larger $b$ values are expected.

A direct estimate of $b$ is obtained from spectrally resolved 
CO $\nu=1-0$ absorption lines toward the north-west nucleus of IRAS\,08572+3915 
\citep*[][]{geballe06}, which shows a complex velocity structure. A low column 
density, cold CO component absorbs near the systemic velocity. 
The CO absorption is however dominated by a broad (FWHM 
$\sim 200\,\mathrm{km\,s}^{-1}$) blue-shifted warm ($\gtrsim$ 200 K) 
gas component.
This is most likely the conglomerate of a velocity distribution over
multiple lines of sight within the beam.
The continuum confusion then requires a higher column density than 
that of $2\times 10^{18}$ cm$^{-2}$ estimated from the observed 
optical depth by \citet*[][]{geballe06}.
We have fitted synthetic LTE absorption spectra to the higher 
excitation CO lines using the spectrum of \citet[][]{geballe06} 
but with the spectral resolution reduced to match the observed profile
width ($R\sim1500$).
This allows to estimate a velocity 
averaged column density for a given $b$ value. 
Values $< 10\,\mathrm{km\,s}^{-1}$ result in progressively poorer fits.
Good fits are made for $b$ values of $10-25\,\mathrm{km\,s}^{-1}$ 
requiring column densities of $\sim2\times 10^{19}$ cm$^{-2}$ down to 
$\sim5\times 10^{18}$ cm$^{-2}$.
The Doppler $b$-value has therefore been fixed to $20\,\mathrm{km\,s}^{-1}$ 
for all sources.

Recent observations using TEXES \citep*[][]{lacy02}, visitor instrument 
on Gemini-North, have
revealed spectrally resolved (FWHM $\sim 80\,\mathrm{km\,s}^{-1}$) 
blue-shifted absorption in the C$_2$H$_2$ \hbox{$\nu_5=1-0$} R(13) 
ro-vibrational line toward one of our sources, NGC\,4418 (Knez, private 
communication). The current data do not allow to put more stringent 
constraints on the Doppler $b$-value. However, the spectrally resolved 
C$_2$H$_2$ absorption could be an indication that multiple (spatial and 
velocity) components of the warm and dense gas are common in (U)LIRG nuclei.

\subsection{Fit results}
\label{sec:fits}
\begin{figure}
\plotone{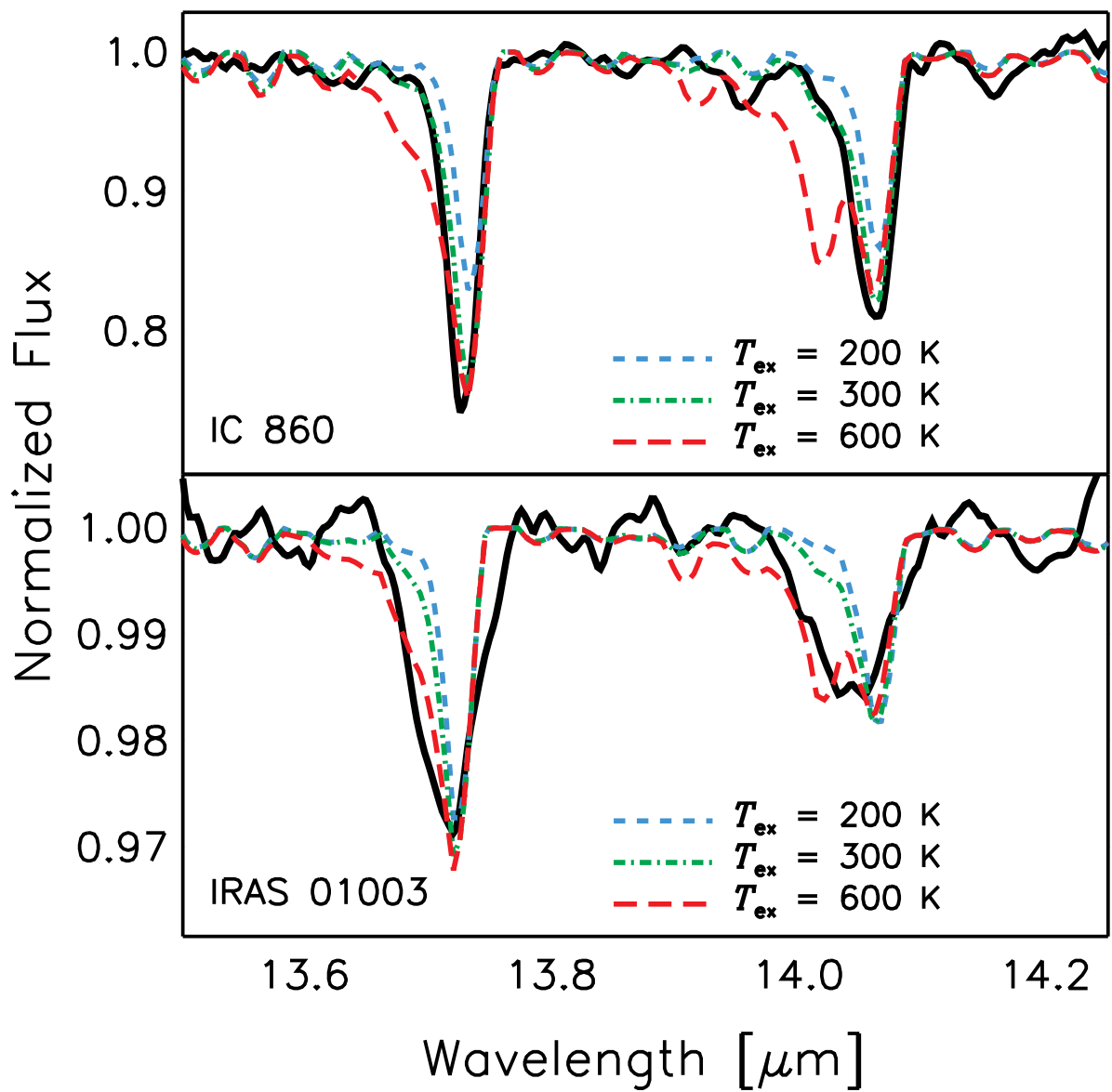}
\caption{\label{fig:mol_fit}%
Illustration of the temperature sensitivity of the C$_2$H$_2$ and 
HCN $Q$-branch profiles in direct comparison to the observed spectra 
of IC\,860 (best fit $T_{\mathrm ex}=280$\,K) and IRAS\,01003-2238
(best fit $T_{\mathrm ex}=630$\,K).
The shape of the $Q$-branch profile determines the derived 
excitation temperatures with an uncertainty of $\sim30\%$. 
A better constrained error 
estimate is difficult to determine at the IRS resolution.
}
\end{figure}

Table \ref{tab:analysis} lists the derived excitation temperatures 
and column densities from best fit synthetic spectra to the continuum 
divided (U)LIRG spectra. Excitation temperatures ranging from 
200 to 700\,K and column densities of a few $\times 10^{15}$ to  
$10^{17}$ cm$^{-2}$ are observed. These results are derived from a 
simultaneous fit to the three absorption bands, in which the excitation 
temperature, set to be the same for all three molecules, is 
constrained by the $Q$-branch profiles of C$_2$H$_2$ and HCN.
Since the analysis is restricted to that of the analysis of the 
resolved $Q$-branches
(see \ref{sec:mol_analysis}), the derived excitation temperatures 
are not well constrained ($\sim 30\%$). However it does clearly
allow to discriminate between molecular gas with warm ($\sim 200-300$\,K) 
and hot ($\sim 500-700$\,K) excitation temperatures.
This is illustrated in Figure \ref{fig:mol_fit} by fits to the
observed absorption profiles of IC\,860 and IRAS\,01003-2238, for
which we derive excitation temperatures of 280 and 630\,K respectively.

The $Q$-branch profiles of C$_2$H$_2$ and HCN reveal significant contributions 
to the absorption from relatively high rotational levels. Hence, given the 
high critical density of these levels, the absorbing gas is likely very 
dense ($n\gtrsim10^7$ cm$^{-3}$). 
Observations of the ro-vibrational transitions of CO in IRAS\,08572+3915 
and IRAS\,00183-7111 indeed imply densities in excess of 
$3\times 10^6$ cm$^{-3}$ \citep*[][]{geballe06,spoon04}.

\subsection{Abundances}
\label{sec:abundances}
The derived column densities are translated into abundances by using a 
total hydrogen column density obtained from the apparent optical depth of the 
$9.7\mu$m silicate absorption band listed in Table \ref{tab:observations}. 
The optical depth is converted assuming:
$$N_\mathrm{H}=\tau_{9.7}(3.5\times 10^{22})\,\mathrm{cm}^{-2},$$
appropriate for the dust in the solar neighbourhood \citep*{roche84,roche85}.

This relation is most appropriate for embedded (extinction dominated) sources 
probing the absorption along the pencil beam line of sight toward the
warm central continuum emitting region.
To illustrate, for the sample of Galactic massive protostars (see 
Section \ref{sec:hot_cores}) the hydrogen column density derived using 
the silicate optical depth estimated from the ISO-SWS archive spectra 
agrees within a factor of two with the hydrogen column density derived 
from CO $\nu=1-0$ measurements \citep{mitchell90,lahuis00}.

Another estimate of the applicability of this method is provided
through the analysis of the 4.6\,$\mu$m CO $\nu=1-0$ absorption spectrum
toward IRAS\,08572+3915NW, which
gives CO column densities of $\sim 5 \times 10^{18}\,\mathrm{cm}^{-2}$
up to $\sim 2\times 10^{19}\,\mathrm{cm}^{-2}$ depending on the 
assumed value of $b$ (see Section \ref{sec:mol_analysis}). 
Adopting $N_\mathrm{CO}/N_\mathrm{H}=10^{-4}$ (all gas-phase carbon in CO)
this results in 
$N_\mathrm{H}=5\times 10^{22} - 2\times 10^{23}\,\mathrm{cm}^{-2}$.
This is in reasonable agreement
with the hydrogen column density derived from the silicate optical depth 
($N_\mathrm{H} \sim 1.5\times 10^{23}\,\mathrm{cm}^{-2}$).
If anything, the IRAS\,08572+3915NW data suggest the hydrogen column density 
might be slightly overestimated (e.g. as a result of foreground extinction) 
resulting in underestimating the derived molecular abundances.
We assume a similar uncertainty for the other sources.

\begin{figure*}
\plotone{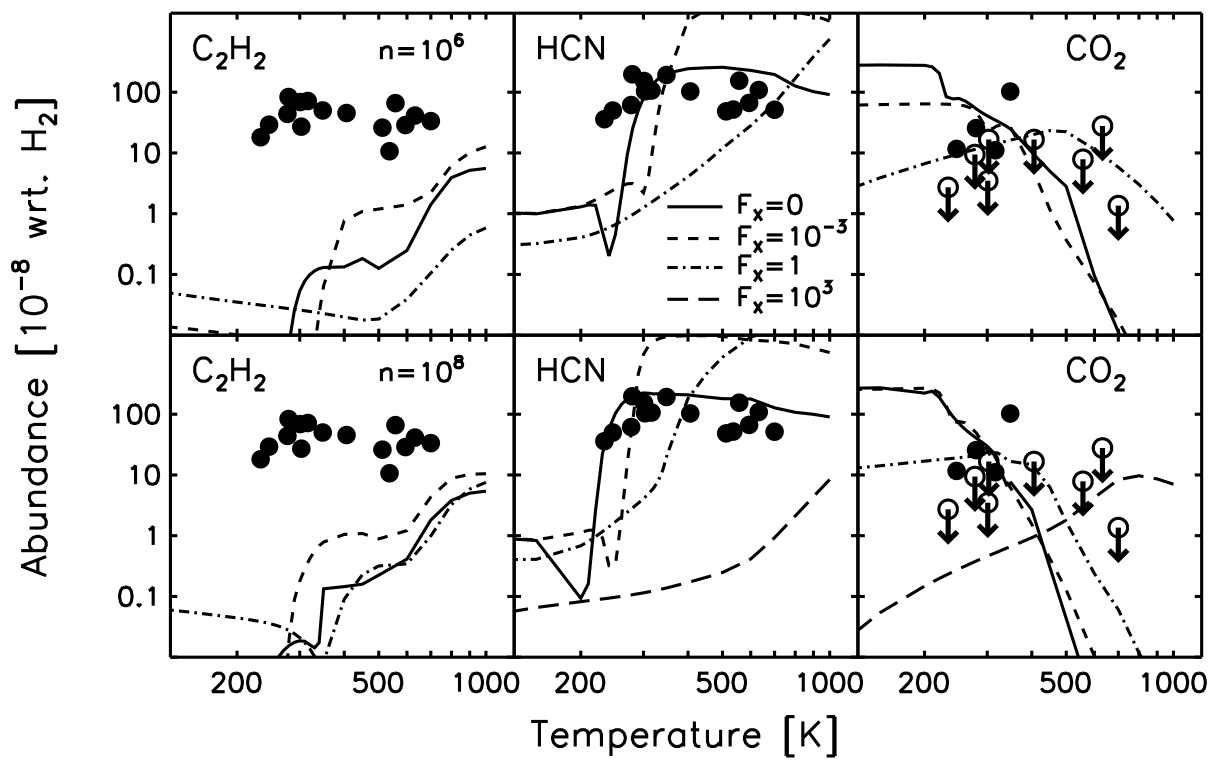}
\caption{\label{fig:chem_model}%
Results of equilibrium X-ray enhanced chemical models for
the three observed molecules.
The calculated abundances are shown for moderate ($10^6\,{\rm cm^{-3}}$)
and high density ($10^8\,{\rm cm^{-3}}$) and four X-ray flux levels 
($F_{\mathrm X}=0$, $10^{-3}$, 1 and $10^3$  erg cm$^{-2}$ s$^{-1}$ plotted with
solid, dashed, dot-dashed, and long-dashed lines). Note that except
for HCN and CO$_2$ at high density, the abundances for the highest
X-ray flux are below the plotted range. 
Overplotted are the observed abundances.
Details of the model can be found in \citet{stauber05,stauber06}.
}
\end{figure*}

The derived abundances and H$_2$ column densities (assuming 
$N_{\mathrm{H}_2}=N_{\mathrm{H}}/2$) are listed in Table \ref{tab:analysis}.
The HCN abundance is up to three orders of magnitude higher 
than in cold molecular clouds \citep*[][]{bergin97}. 
High abundances of HCN, C$_2$H$_2$, and CO$_2$ have also been
measured for Galactic massive protostars (see Section \ref{sec:hot_cores}).
Figures \ref{fig:chem_model} and \ref{fig:abundances}
compare the abundances to chemical models and results from studies 
of Galactic massive young stellar objects (Section \ref{sec:discussion}).

\subsection{Gas temperature}

The derived excitation temperatures range from 200 to 700 K. 
While statistics are small, of the seven sources with the highest gas 
excitation temperatures ($T_{\mathrm ex}=400-700$\,K) five sources show 
the highest 5.5 $\mu$m to 28 $\mu$m continuum flux ratio.
These galaxies also show a rising near-IR continuum typical for hot dust
while the other galaxies show a decreasing near-IR continuum characteristic
for emission dominated by stellar photospheres. It seems thus that there is a
connection between the observed dust and molecular excitation temperatures.

The presence of absorption (rather than emission) lines implies a strong
gradient in the temperature and an underlying hot-dust continuum
against which the absorption lines are formed.
However, the derived excitation temperatures of the molecular gas
for the warmest sources ($T_{\mathrm ex} =400-700$\,K) are higher by a factor
of 2 to 3 than the mid-IR color temperature derived from the 
5.5 to 28.0\,$\mu$m flux ratio.
While the discrepancy between the excitation temperatures and the 
mid-IR color temperature may (in part) reflect the uncertainties in the 
analysis, the observed colder dust continuum may point towards the
importance of continuum extinction by
(a shell of) colder foreground dust exterior to the C$_2$H$_2$ and HCN 
absorbing layer, or the predominance of emission of cold nuclear dust
within the IRS aperture (i.e. a small beam filling factor for the warm 
dust and gas). Both may result in an added uncertainty
in the derived abundances resulting from overestimating the hydrogen column
density or from a covering factor of the molecule rich warm
gas less than unity (see Section \ref{sec:mol_analysis}).

\section{Discussion}
\label{sec:discussion}

\subsection{Warm molecular gas in (U)LIRGs}
\label{subsec:warm-mol-gas}
The {\it Spitzer}-IRS spectra reported here reveal surprisingly strong mid-IR 
absorption bands of C$_2$H$_2$, HCN and CO$_2$ toward a sample of (U)LIRG 
nuclei. These absorptions reveal the presence of copious amounts of warm 
molecular material and are in line with earlier detections of warm 
CO gas \citep*[][]{spoon04,spoon06a,geballe06}. This molecular gas is 
very dense ($n\gtrsim 3\times 10^6$ cm$^{-3}$) and warm ($T\sim 200-700$ K). 
Given the column density and estimated density, this gas occupies only 
a small fraction of the nuclear region ($\sim 0.01$ pc) and, given 
its high temperature, is likely located near the intrinsic mid-infrared 
source of luminosity of these regions.

At present there are still substantial uncertainties in accurately inferring
the nature of the power source dominating the bolometric luminosity in
infrared luminous galaxies. As discussed in \citet{armus06b},  it appears
that there are disagreements among various diagnostic diagrams based on line
ratios, continuum colors or PAH equivalent width.
Perusal of the $2-40\mu$m spectra 
reveal that the sources with  molecular gas absorptions  appear to be a 
``mixed bag''. While most sources show a deep silicate feature, some 
(e.g., Mrk231, IRAS\,01003-2238, IRAS\,05189-2524 and IRAS\,13218+0552) do not.
Indeed, the $2-40\mu$m continuum spectra of these sources show strong 
similarities with the spectra of bona fide AGN-dominated spectra with 
shallow silicate features.
In contrast, sources like IRAS\,17208-0014 and IRAS\,22491-1808 look 
quite starburst-like. Finally, sources such as IRAS\,08572+3915 show 
neither signs of starburst activity
(e.g., PAH features) nor signs of AGN activity (e.g., high ionization lines 
or a broad H$\alpha$ line). While the presence of hot dust,
as visible in the near-IR, may be taken as evidence for the presence of
an embedded AGN in this source, any deeply obscured source -- including 
deeply embedded protostars -- will produce hot dust. Indeed, the overall 
mid-IR spectrum of IRAS\,08572+3915 resembles that of many galactic
massive protostars, such as AFGL\,2591, with a preponderance of dust 
absorption features.

The C$_2$H$_2$, HCN, and CO$_2$ molecular lines may provide a new probe of 
the conditions in the inner deeply obscured nuclei and hence shed light 
on the ultimate power source of (U)LIRGs; e.g., starburst phenomena 
versus AGN activity. 
In particular, the presence of copious amounts of warm molecular gas enriched 
in C$_2$H$_2$, HCN, and CO$_2$ is common in regions of massive star 
formation \citep{boonman03,lahuis00,lacy89,evans91}, while, on the 
other hand, X-rays associated with a central black hole can influence
molecular chemistry and abundances far beyond the mean free path
length of stellar (E)UV photons.
The molecular observations are discussed in the light of these scenarios in
the next subsections. While the possible errors associated with the 
derived abundances are recognized, the discussion will center 
on Figures \ref{fig:chem_model} and \ref{fig:abundances}.

\subsection{AGN activity and X-ray driven chemistry}
\label{sec:ss_model}

Sub-millimeter studies have revealed high HCN/CO (and HCN/HCO$^+$) 
abundance ratios in Seyfert type I nuclei and this has been taken as 
an indication of the importance of X-ray illumination for the composition 
of the molecular gas in these nuclei \citep*[][]{imanishi06,kohno05}. 
Some of the sources in our sample are known to harbor an AGN. 
In particular, Mrk 231 and IRAS 13218+0552 are QSOs, IRAS 08572+3915 and 
UGC\,5101 are LINERs, and IRAS\,01003-2238, IRAS\,05189-2524 and
Mrk\,273 have optical LINER/Seyfert\,2 spectra
(see Table \ref{tab:observations}).
Hence, the observed high abundance of HCN, and by inference 
C$_2$H$_2$ (which has no pure rotational transitions), may indicate 
the presence of a buried AGN in all of our sources. The complex geometry 
and inhomogeneous dust distribution associated with the central toroid 
may then preclude our detection of the putative X-rays photons from the 
central engine driving the chemistry of the molecular gas. The mid-IR 
observations presented in this paper add to the submillimeter results 
the predominance of {\it warm} molecular gas. Within the AGN heating scenario, 
it seems obvious to attribute the high temperature of the molecular gas to the 
importance of X-ray heating in a so-called XDR \citep*{maloney96,meijerink05}.
Alternatively \citet{gao04} argue that most (U)LIRGs are dominated by 
starburst power rather than AGN power based on the HCN $J=1-0$ luminosity and
CO $J=1-0$ luminosity relation. In a starburst scenario the HCN/CO ratio 
traces the amount of dense molecular gas in the galactic ISM.
Our observations may present an additional test for these two scenarios.

X-rays are not only an efficient heating agent of the gas but also have a major 
influence on the chemical composition. This has been explored in the 
context of AGNs by \citet*{maloney96,meijerink05}, and for regions around 
protostars by \citet*{stauber05,stauber06}.  At the high densities 
relevant for these 
regions, the gas temperature is set by a balance between X-ray heating and 
gas-dust collisional cooling. As an example, a gas temperature of 200 K at a 
density of $\sim 3\times 10^6$ cm$^{-3}$ requires a X-ray flux of 
$\sim 30$ erg cm$^{-2}$ s$^{-1}$ which for a typical Seyfert galaxy with 
an X-ray luminosity of $10^{44}$ erg s$^{-1}$ corresponds to a distance 
of 50 pc for an unobscured view of the central engine. 
For the molecular features to be seen in absorption, the mid-IR continuum 
has to arise from an inner, warmer zone (e.g., exposed to a higher X-ray flux).
Since most of these (U)LIRGs show no evidence for strong X-ray emission while 
the column density associated with the molecular gas 
($10^{22}-10^{23}$ cm$^{-2}$) 
is small compared to X-ray attenuation lengths, in this AGN-scenario, the 
direct view of the nucleus would have to be blocked by a high-column-density, 
edge-on toroid while the warm molecular gas is offset to the polar regions 
and has a little-obstructed view of the nucleus.

An AGN origin for the high abundance of warm HCN, C$_2$H$_2$, and CO$_2$ in 
these (U)LIRGs however faces severe theoretical difficulties. Specifically, 
the high X-ray flux required to explain the observed temperatures readily 
dissociates the molecules 
and drives the gas atomic \citep*{maloney96,meijerink05}. Indeed, calculated 
HCN abundances are typically less than $10^{-8}$ wrt. H in X-ray 
illuminated gas while C$_2$H$_2$ is virtually destroyed for X-ray fluxes
in excess of 1 erg cm$^{-2}$ s$^{-1}$. 
This is illustrated in Figure \ref{fig:chem_model} which presents the 
static steady-state chemical composition after equilibrium is reached 
($\sim 10^5$\,yr) for clouds illuminated by X-rays. Here, static means 
that the chemistry has been 
evolved in time -- while the X-ray flux, temperature, and density are fixed -- 
until steady state is achieved in the chemical abundances 
\citep*[see also][]{stauber05,stauber06}. The X-ray spectrum is assumed to be
thermal in nature with $T_\mathrm{X}$=$3\times10^7$\,K. 
The chemistry is relatively insensitive to the assumed shape of the 
spectrum \citep[e.g.,][]{maloney96}. 
As these models demonstrate, strong X-ray irradiation leads to decreased 
abundances of molecular species such as HCN and C$_2$H$_2$.  
These results are consistent with earlier studies \citep*[e.g.][]{lepp96} which 
show an initial increase in the HCN abundance with increasing ionization rate 
($\xi/n_\mathrm{H}$ up to $10^{-18}\,\mathrm{cm}^{-3}\,\mathrm{s}^{-1}$)
but then an effectively complete destruction of HCN for 
$\xi/n_\mathrm{H}\gtrsim10^{-17}\,\mathrm{cm}^{-3}\,\mathrm{s}^{-1}$. 
For comparison, the X-ray flux implied by a gas temperature of 200 K 
($\simeq$ 30 erg cm$^{-2}$ s$^{-1}$) corresponds to an ionization rate of 
$5\times 10^{-17}$ cm$^{-3}$ s$^{-1}$ at a density of $10^6$ cm$^{-3}$.
Clearly, the warm molecular gas revealed by these observations is inconsistent 
with strong X-ray illumination.

\subsection{Static hot-core chemistry}
\label{sec:hot_cores}

At first sight our mid-infrared results resemble those toward a 
sample of massive Galactic young stellar objects (YSOs). For these, 
the molecular absorption features 
originate in the warm, dense gas near the newly formed star during 
the so-called hot core phase \citep{boonman03,lahuis00,lacy89,evans91}.
This is generally thought to represent a very early phase 
($\Delta t\leq$10$^5$\,yr) in the formation of massive stars, during 
the transition from the deeply embedded stage to the phase where 
ionizing photons can escape the protostellar accretion envelope 
and create first a hyper- and later an ultra-compact H{\sc ii} region 
\citep*[see review by][]{cesaroni05}. The molecular composition of 
the hot core is very different from that of cold molecular clouds. 
This is thought to reflect the evaporation of ices -- resulting from 
accretion and surface chemistry during the cold preceding dark cloud phase --
when the envelope is heated  by the newly formed star \citep*{walmsley93}.  
Subsequent high temperature gas-phase chemistry 
significantly enhances the abundance of e.g. CH$_4$, C$_2$H$_2$ and HCN 
up to three orders of magnitude, compared to cold dense molecular cloud 
abundances, for the most evolved sources \citep{viti99,doty02,rodgers03}.

\begin{figure*}[t]
\plotone{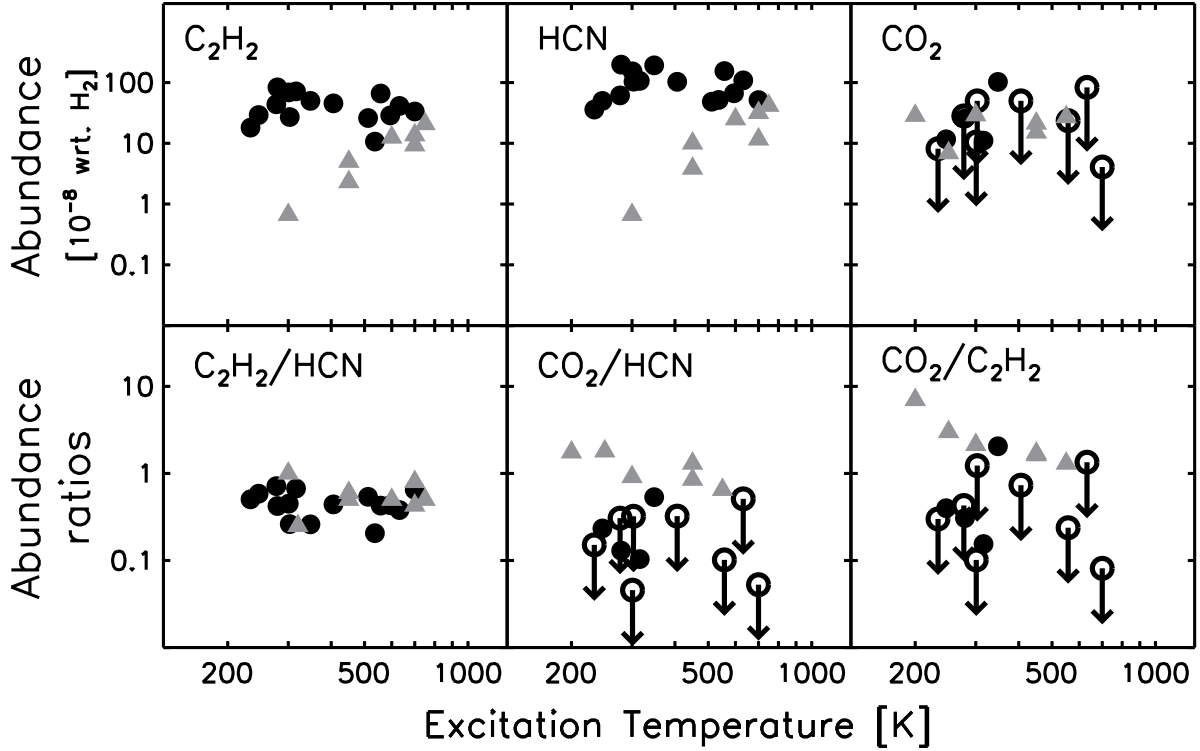}
\caption{\label{fig:abundances}%
Top: observed C$_2$H$_2$, HCN and CO$_2$ abundances using a total 
H$_2$ column obtained from the $9.8\,\mu$m silicate absorption band.
Bottom: abundance ratios of C$_2$H$_2$, HCN and CO$_2$.
Both the absolute abundances and the abundance ratios are indicated with
circles and are presented 
as functions of excitation temperature. Included are the observed values 
toward a sample of Galactic massive YSOs \citep{boonman03,lahuis00}
plotted with grey triangle symbols.
The derived absolute abundances toward the (U)LIRGs show significant 
enhancements in the abundances of C$_2$H$_2$ and HCN for sources with 
excitation temperatures below 500\,K. The C$_2$H$_2$/HCN abundance 
ratios are quite similar, however the CO$_2$/C$_2$H$_2$ and 
CO$_2$/HCN ratios toward the (U)LIRGs are systematically lower
than toward the Galactic massive YSOs suggesting that a different 
chemistry may apply.
}
\end{figure*}

In Figure \ref{fig:abundances} the observed abundances and abundance ratios 
toward the (U)LIRGs are compared to those of a sample of Galactic massive YSOs. 
Both the YSO and the (U)LIRG sample show comparable, large variations in 
gas temperature and abundances. However, the HCN and C$_2$H$_2$ abundances
show a positive correlation with the gas temperature for the YSO
sample but not for the (U)LIRG sample. Indeed, at low temperatures 
($<400$ K), the (U)LIRG abundances are some two orders of magnitude larger 
than those in YSO spectra.

For the Galactic YSOs, the observed correlation between the excitation 
temperature and the observed abundances (Figure~\ref{fig:abundances}) 
are in fair agreement with chemical models for such regions \citep*{stauber05}.
The absence of a similar temperature correlation for (U)LIRGs may reflect a 
systematic (and large) error in the abundance determination 
(Section \ref{sec:abundances})  or it may reflect a true difference 
between the chemical or physical evolution of the warm gas of the YSOs 
and (U)LIRGs. In this respect, we note that the C$_2$H$_2$/HCN ratio for 
both samples are very similar, however the CO$_2$/HCN and CO$_2$/C$_2$H$_2$ 
ratios are reduced for the (U)LIRG sample up to an order of magnitude. 
If static steady-state hot-core chemistry were to apply to the (U)LIRGs 
as it does to the YSOs, CO$_2$ would have been detected toward most of the 
sources in our sample. The fact it has not, suggests that a different 
chemistry may apply to the warm molecular gas in these (U)LIRGs.

\subsection{Pressure confined starburst chemistry}
\label{sec:outflow}
Static steady-state hot-core and X-ray chemical models have difficulties 
to consistently reproduce the observed high abundances of warm C$_2$H$_2$ 
and HCN gas and low abundance of CO$_2$ gas for most of our (U)LIRG sample. 
Essentially, static chemistry has a hard time producing high abundance of 
HCN and C$_2$H$_2$ in the warm ($100-400$\,K) gas characteristic 
for most of our (U)LIRG sample, while at the same time reducing
the CO$_2$ abundance. One possible solution to this conundrum 
is to produce high C$_2$H$_2$ and HCN, and low CO$_2$ abundances 
in hot gas ($\gtrsim 800$\,K), and then transport it outward to the cooler 
gas on a dynamical timescale which is rapid compared to the chemical timescale. 
At high temperatures, the hydrocarbon and nitrogen chemistry are enhanced
as most of the oxygen is converted into water by neutral-neutral
reactions. The abundances of molecules such as C$_{2}$H$_{2}$,
CH$_{4}$, and HCN can be increased by orders of magnitude
\citep*[e.g.][]{doty02,rodgers03} while at the same time the
formation of CO$_{2}$ is reduced as its primary formation route
through OH is blocked. 
For cold Galactic molecular clouds, the chemical timescale is set by the 
cosmic ray ionization rate and is about $3\times 10^5$ yr, independent 
of density \citep{glassgold73}. 
In a (U)LIRG (or AGN) environment, the  cosmic ray flux may be increased 
considerably, shortening this timescale. Likewise, neutral-neutral reaction 
channels may open up in warm gas, further reducing the chemical timescale.
In any case, these timescales are much shorter than the evolutionary 
timescale of massive stars or of the starburst. With the typical sizescale 
of the warm molecular zone  ($\sim 3\times 10^{16}$ cm), a timescale of 
$3\times 10^5$ yr translates into a `diffusion' velocity of only 
0.03 km\,s$^{-1}$. 
Even a 100 times faster chemical timescale only requires a `diffusion' 
velocity of a few km\,s$^{-1}$. Since the chemical models are well capable of 
explaining high abundances for both C$_2$H$_2$ and HCN and a low abundance
for CO$_2$ at high temperatures and densities, and in
light of the discussion above, this `mixing' is attributed to the global 
activity created by pressure confined massive starburst activity. 
The strong gravitational potential in the nuclei of these galaxies
inhibits the disruption of the 
surrounding warm molecular envelopes by H{\sc ii} regions and supernovae, 
while producing sufficient turbulent (or wind) motion to distribute warm  
C$_2$H$_2$- and HCN-rich and CO$_2$-poor gas in the colder outer 
envelope regions.

\section{Summary}

We have observed the absorption lines due to the $Q$-branch transitions of 
C$_2$H$_2$, HCN, and CO$_2$ in the mid-IR spectra of a large number of (U)LIRGs. 
These observations reveal for the first time the presence of copious amounts 
of warm (200-700 K), dense ($n>10^7$ cm$^{-3}$) molecular gas in these nuclei. 
The origin of this warm molecular gas is unclear. Theoretical models show that 
the X-ray fluxes implied by the elevated gas temperatures rapidly destroy 
these molecules and hence it is unlikely that this warm molecular 
gas is associated with a dense toroid surrounding an active central engine. 
Warm, dense gas rich in C$_2$H$_2$, HCN, and CO$_2$ is commonly observed 
towards galactic massive protostars and is associated with a short-lived 
phase -- the Hot Core phase -- before advancing ionization fronts disrupt the core. 
The high molecular abundances in galactic Hot Cores are well understood 
chemically. However, the derived abundances of C$_2$H$_2$ and HCN for the 
cooler (U)LIRGs ($T_\mathrm{ex} \simeq 200-400$ K) in our sample 
as well as the C$_2$H$_2$/CO$_2$ 
and HCN/CO$_2$ ratios are very different from those in galactic protostars. 
We suggest that this warm molecular gas is associated with a phase of deeply 
embedded star formation in (U)LIRGs where the high pressures and densities have
inhibited the disruption of the star forming, molecular cores 
by prohibiting the expansion of HII regions, trapping the star formation
process in an extended Hot-Core phase. 
The chemical differences between 
these (U)LIRGs and galactic Hot Cores may then reflect enhanced mixing 
between warm and cold phases due to the high degree of 
turbulence or wind motion associated with extreme starburst environments.  
Pressure confined massive starburst activity may thus be the driving
energy source behind the observed C$_2$H$_2$ and HCN rich warm molecular gas
and responsible for most of the near-IR characteristics of the deeply
obscured (U)LIRGs.

The current analysis is predominately based on the observed abundances
derived from moderate resolution, mid-IR vibration-rotation absorption bands 
of C$_2$H$_2$ and HCN. This analysis can be extended
using high resolution, velocity resolved, ground-based studies of the 
$P$- and $R$ branches of these molecules as well as from 
the $\nu=1-0$ rotation-vibration band of CO. In addition, observations 
of more molecular rotation-vibration bands and high excitation 
submillimeter lines of a large sample of molecules may be instrumental for 
progress in this field.  Specifically, velocity and spatially resolved
infrared and millimeter data can constrain the source morphology.
In combination with optimized physicochemical models for individual
sources it may become possible to draw firm conclusions about the physical 
characteristics of the warm molecular gas and the true
nature of the power sources in the (U)LIRG nuclei.

\acknowledgments
The authors would like to thank Tom Geballe for sharing early results 
and the CO data of IRAS\,08572+3915NW; John Lacy and Claudia Knez for
sharing the TEXES data; and Bernhard Brandl, Kees Dullemond, 
Masa Imanishi, David Rupke, and Marco Spaans for many useful discussions.
Astrochemistry in Leiden is supported by a Spinoza grant from NWO.

%
%
\clearpage
\begin{deluxetable}{lccrrrcccc}
\rotate
\tabletypesize{\scriptsize}
\tablecaption{Observation details and basic source properties\label{tab:observations}}
\tablehead{\colhead{Target} & 
           \colhead{AOR Key} &
           \colhead{Pipeline} &
           \colhead{Observed} &
           \colhead{Integration Time\tablenotemark{a}} &
           \colhead{$z$} & 
           \colhead{log $L_{\rm IR}$} & 
           \colhead{$\tau_{9.7}$} &
           \colhead{Optical/Near-IR Class\tablenotemark{b}} &
           \colhead{$D_L$\tablenotemark{c}} 
       \\
           \colhead{} &
           \colhead{} &
           \colhead{} &
           \colhead{} &
           \colhead{(sec)} &
           \colhead{} &
           \colhead{($L_\odot$)} &
           \colhead{} &
           \colhead{} &
           \colhead{(Mpc)} 
       }
\startdata
17208-0014 & 4986624 & S12.0.2 & 27 Mar 2004 & $ 31\times6\times2$ & 0.0430 & 12.46 & 1.9 & H\,{\sc ii}       &  188 \\ 
Arp\,220   & 4983808 & S12.0.2 & 29 Feb 2004 & $ 31\times6\times2$ & 0.0181 & 12.17 & 3.3 & LINER$^1$         &   78 \\ 
IC\,860    & 6652416 & S13.2.0 & 10 Feb 2005 & $121\times2\times2$ & 0.0112 & 10.96 & 2.1 &                   &   48 \\ 
22491-1808 & 4990976 & S13.2.0 & 24 Jun 2004 & $121\times2\times2$ & 0.0773 & 12.19 & 1.1 & Starburst$^{2,3}$ &  346 \\ 
NGC\,4418  & 4935168 & S13.2.0 &  8 Jul 2005 & $ 31\times6\times2$ & 0.0073 & 11.03 & 4.4 &                   &   31 \\ 
13218+0552 & 4979200 & S13.2.0 & 17 Jul 2004 & $121\times3\times2$ & 0.2051 & 12.71 & 0.8 & QSO/Sey-1$^{4,5}$ &  998 \\ 
15250+3609 & 4983040 & S12.0.2 &  4 Mar 2004 & $ 31\times6\times2$ & 0.0554 & 12.05 & 3.8 &                   &  244 \\ 
05189-2524 & 4969216 & S13.2.0 & 22 Mar 2004 & $ 31\times6\times2$ & 0.0426 & 12.19 & 0.4 & Seyfert-2/1$^6$   &  186 \\
Mrk\,231   & 4978688 & S13.2.0 & 14 Apr 2004 & $ 31\times6\times2$ & 0.0422 & 12.52 & 0.8 & Seyfert-1$^2$     &  184 \\ 
Mrk\,273   & 4980224 & S12.0.2 & 14 Apr 2004 & $ 31\times6\times2$ & 0.0378 & 12.15 & 1.9 & Seyfert-2         &  164 \\
00397-1312 & 4963584 & S12.0.2 &  4 Jan 2004 & $121\times3\times2$ & 0.2617 & 12.90 & 3.3 &                   & 1317 \\ 
20100-4156 & 4989696 & S13.2.0 & 13 Apr 2004 & $121\times2\times2$ & 0.1296 & 12.65 & 3.3 &                   &  601 \\ 
UGC\,5101  & 4973056 & S13.2.0 & 23 Mar 2004 & $ 31\times6\times2$ & 0.0400 & 12.00 & 1.7 & LINER$^3$         &  164 \\
01003-2238 & 4972032 & S12.0.2 &  4 Jan 2004 & $121\times2\times2$ & 0.1177 & 12.29 & 0.7 & Seyfert-2$^1$     &  542 \\ 
08572+3915 & 4972032 & S13.2.0 & 15 Apr 2004 & $ 31\times6\times2$ & 0.0584 & 12.10 & 4.2 & LINER$^1$         &  258 \\ 
  &   &   &   &   &  &   \\
02530+0211 & 6652160 & S13.2.0 & 10 Feb 2005 & $ 31\times3\times2$ & 0.0276 & 11.04 & 3.7 &                   &  119 \\ 
12112+0305 & 4977664 & S13.2.0 &  4 Jan 2004 & $121\times2\times2$ & 0.0727 & 12.33 & 1.3 & LINER$^1$         &  324 \\
14348-1447 & 4981248 & S13.2.0 &  7 Feb 2004 & $121\times2\times2$ & 0.0827 & 12.35 & 2.1 & LINER$^2$         &  372 \\
NGC\,6240  & 4985600 & S13.2.0 &  4 Mar 2004 & $ 31\times6\times2$ & 0.0245 & 11.84 & 1.2 & LINER             &  119 \\
\enddata
\tablenotetext{a}{Ramp integration time $\times$ number of cycles $\times$ number of slit positions}
\tablenotetext{b}{Optical/NIR spectral classifications taken from: 
                  (1) \citet{armus89},
                  (2) \citet{sanders88b},
                  (3) \citet{veilleux95},
                  (4) \citet{low88},
                  (5) \citet{darling02},
                  (6) \citet{veilleux97}
                 }
\tablenotetext{c}{assuming ${\rm H_0=71\,km\,s^{-1}\,Mpc^{-1},\Omega_M=0.27, \Omega_\Lambda=0.73, \Omega_K=0}$}
\end{deluxetable}



%

\clearpage
\begin{deluxetable}{lcccccccccc}
\rotate
\tabletypesize{\scriptsize}
\tablecaption{LTE excitation parameters and spectral characteristics\label{tab:analysis}}
\tablehead{\colhead{Target} & 
           \colhead{$T_{\rm ex}$\tablenotemark{a}} &
           \colhead{$N_{\rm C_2H_2}$\tablenotemark{a}} &
           \colhead{$N_{\rm HCN}$\tablenotemark{a}} &
           \colhead{$N_{\rm CO_2}$\tablenotemark{a}} & 
           \colhead{$N_{\rm H_2}$} & 
           \colhead{$x_{\rm C_2H_2}$\tablenotemark{b}} &
           \colhead{$x_{\rm HCN}$\tablenotemark{b}} &
           \colhead{$x_{\rm CO_2}$\tablenotemark{b}} & 
           \colhead{cont\tablenotemark{c}} &
           \colhead{SNR\tablenotemark{d}}
       \\
           \colhead{} &
           \colhead{(K)} &
           \colhead{($10^{16}{\rm cm^{-2}}$)} &
           \colhead{($10^{16}{\rm cm^{-2}}$)} &
           \colhead{($10^{16}{\rm cm^{-2}}$)} &
           \colhead{($10^{22}{\rm cm^{-2}}$)} &
           \colhead{($10^{-8}$)} &
           \colhead{($10^{-8}$)} &
           \colhead{($10^{-8}$)} &
           \colhead{(Jy)} &
           \colhead{}
       }
\startdata
17208-0014  & 230 & 0.6 & 1.2 & $<0.2$ &  3.3  & 18  &  36  & $<6$  & 0.20 & 200 \\
Arp\,220    & 250 & 1.7 & 2.9 & 0.7    &  5.8  & 29  &  50  &  12   & 0.93 & 150 \\
IC\,860     & 280 & 3.1 & 7.2 & 0.9    &  3.7  & 84  & 195  &  24   & 0.09 & 150 \\
22491-1808  & 280 & 1.0 & 1.4 & $<0.3$ &  1.9  & 53  &  74  & $<16$ & 0.07 & 130 \\
NGC\,4418   & 300 & 5.3 & 12. & $<0.4$ &  7.7  & 69  & 156  & $<5$  & 2.34 & 150 \\
13218+0552  & 300 & 0.4 & 1.5 & $<0.5$ &  1.4  & 28  & 107  & $<36$ & 0.26 & 180 \\
15250+3609  & 320 & 4.7 & 7.0 & 0.7    &  6.7  & 70  & 105  &  10   & 0.27 &  90 \\
05189-2524  & 350 & 0.4 & 1.5 & 0.8    &  0.7  & 57  & 214  & 114   & 1.05 & 200 \\
Mrk\,231    & 410 & 0.6 & 1.4 & $<0.4$ &  1.4  & 43  & 100  & $<29$ & 2.83 & 200 \\
Mrk\,273    & 510 & 0.9 & 1.7 & $<0.4$ &  3.3  & 27  &  52  & $<12$ & 0.37 & 200 \\
00397-1312  & 540 & 0.6 & 3.0 & $<0.4$ &  5.8  & 10  &  52  & $<7$  & 0.12 & 150 \\
20100-4156  & 560 & 3.8 & 8.9 & $<0.9$ &  5.8  & 66  & 153  & $<15$ & 0.11 &  70 \\
UGC\,5101   & 590 & 1.0 & 2.2 & $<0.5$ &  3.0  & 33  &  73  & $<17$ & 0.24 & 130 \\
01003-2238  & 630 & 0.5 & 1.3 & $<0.8$ &  1.2  & 42  & 108  & $<67$ & 0.27 & 200 \\
08572+3915  & 700 & 2.4 & 3.7 & $<0.5$ &  7.4  & 32  &  50  & $<7$  & 0.74 & 150 \\
            &     &     &     &        &       &     &      &       &      &     \\
02530+0211  & 300\tablenotemark{e} & $<0.5$ & $<3  $ & $<0.3$ & 6.5  &  $<8$  &  $<46$  & $<5$ & 0.25 & 100 \\
12112+0305  & 300\tablenotemark{e} & $<0.8$ & $<2.3$ & $<0.6$ & 3.0  &  $<27$ &  $<77$  & $<20$ & 0.07 & 150 \\
14348-1447  & 300\tablenotemark{e} & $<0.5$ & $<1  $ & $<0.2$ & 3.7  &  $<14$ &  $<27$  & $<5$  & 0.07 &  80 \\
NGC\,6240   & 300\tablenotemark{e} & $<0.2$ & $<0.7$ & $<0.6$ & 2.1  &  $<10$  & $<33$  & $<29$ & 0.68 & 100 \\
\enddata
\tablenotetext{a}{The excitation temperature is poorly constrained and 
                  can be uncertain up to 30\%. 
                  See Section \ref{sec:fits} and Figure \ref{fig:mol_fit} 
                  for details.
                  }
\tablenotetext{b}{Abundances wrt. H$_2$ assuming 
                  $N_\mathrm{H} = \tau_{9.7} ( 3.5 \times 10^{22})\,\mathrm{cm}^{-2}$
                  and $N_\mathrm{H} = 2\times N_{\mathrm{H}_2}$ 
                  (see Section \ref{sec:abundances}).}
\tablenotetext{c}{Continuum at 14 $\mu$m in the rest wavelength frame.}
\tablenotetext{d}{Signal to Noise Ratio (SNR) estimated from the residuals after subtraction 
                  of the synthetic spectrum. It varies over the covered 
                  wavelength range.
                 }
\tablenotetext{e}{Temperature fixed to 300\,K for derivation of 
                  column density upper limit estimates.}
\end{deluxetable}
%

\clearpage

\clearpage

\clearpage

\clearpage

\end{document}